\documentclass{moriond}
\usepackage{float}
\usepackage{amsmath}
\usepackage{graphicx}

\def\Journal#1#2#3#4{{#1} {\bf #2}, #3 (#4)}

\def\PRD{{\em Phys. Rev.} D}

\def\be{\begin{equation}}
\def\ee{\end{equation}}
\def\bea{\begin{eqnarray}}
\def\eea{\end{eqnarray}}

\newcommand{\Photo}{\includegraphics[height=35mm]{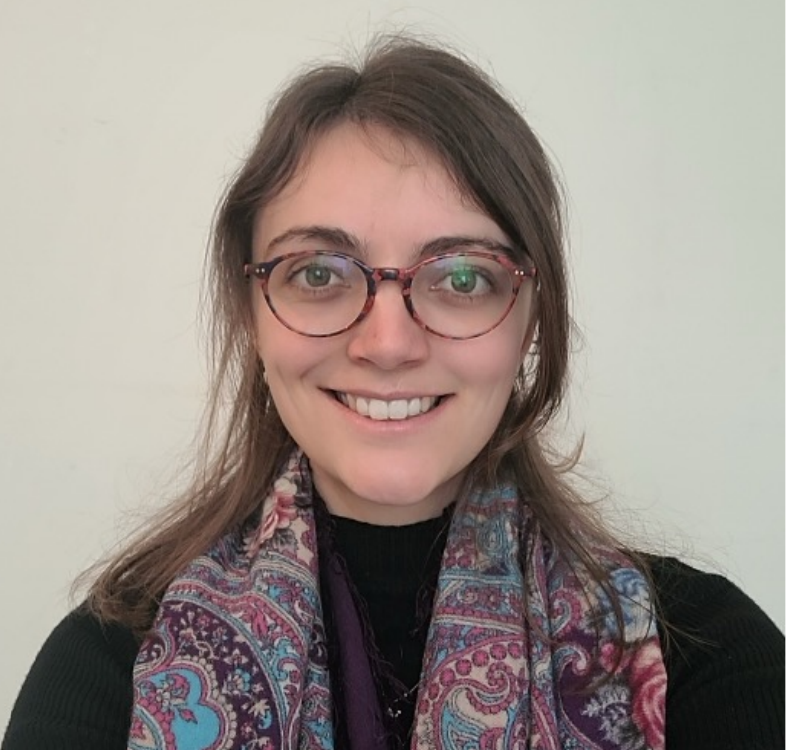}}

\begin{document}

\vspace*{4cm}

\title{Characterizing the correlation properties of the atmospheric emission in the 10-20 GHz range with QUIJOTE MFI data}

\author{Apolline Chappard $^{1,2}$, José Alberto Rubiño-Martín $^{2,3}$,  and Ricardo Tanausú Génova-Santos $^{2,3}$}

\address{ \scriptsize{$^1$ Institut d'Astrophysique Spatiale (IAS), CNRS, Bât 120 – 121 Univ. Paris-Saclay, ORSAY CEDEX, 91405, France \\ $^2$ Instituto de Astrofísica de Canarias (IAC), C/ Vía Láctea, La Laguna, E-38205, Tenerife, Spain \\ $^3$ Departamento de Astrofísica, Universidad de La Laguna (ULL), La Laguna, E-38206, Tenerife, Spain}}

\maketitle

\abstracts{The QUIJOTE MFI instrument (2012-2018) observed the sky at four frequency
bands, namely 11, 13, 17 and 19\,GHz, at 1 degree angular resolution. Using around 10000\,h of observations in the so-called nominal mode, QUIJOTE MFI produced sky maps covering approximately 29000\,deg$^2$. 
Here we use the full database of MFI wide survey observations to characterize the correlation properties of the atmospheric signal in those frequency bands. This information will be
useful to improve the current sky models at these frequencies, and could be
used in further MFI reanalyses, or for the preparation of future observations at
these frequencies (e.g., MFI2 and the Tenerife Microwave Spectrometer). }

\section{Introduction}

The next goal for Cosmic Microwave Background (CMB) research is the detection of primordial B-modes, which are a prediction of the theory of inflation \cite{Ka}. In this theory, at a time of $10^{-36}$\,s after the Big Bang, the size of the universe expanded exponentially by at least 50 to 60\,e-fold in a very brief period of time \cite{Gu}. This process would produce a background of gravitational waves that could be indirectly observed today through their signature on the CMB polarization as B-mode polarization, parametrized with the tensor-to-scalar ratio $r$ \cite{Za}. However, specific models of inflation, characterized by specific inflationary potentials, do predict $r$ to be in certain ranges. For instance, the Starobinsky $R^2$ model predicts $r$ to be between $0.003$ and $0.01$ (depending on the exact value of $n_S$). Today, we only have an upper limit for $r$, being $r<0.030$ \cite{Ga}. Hence, achieving increased sensitivity is imperative to detect smaller signals obscured by various B-mode sources (galactic and extra-galactic), instrumental noise, and primarily, by the atmosphere for ground-based telescopes. In this study, we performed a statistical analysis of part of the data collected by the Multi Frequency Instrument (MFI) of the QUIJOTE (Q-U-I JOint TEnerife) experiment, the so called MFI wide survey \cite{Ru,Ge}. Our aim was to glean insights into the atmospheric signal, a crucial endeavor for the advancement of next-generation ground-based telescopes, which need a deeper understanding of the impact of the atmosphere.

\section{The QUIJOTE CMB experiment}

\begin{figure}[t!]
    \centering
    \includegraphics[width=0.7\textwidth]{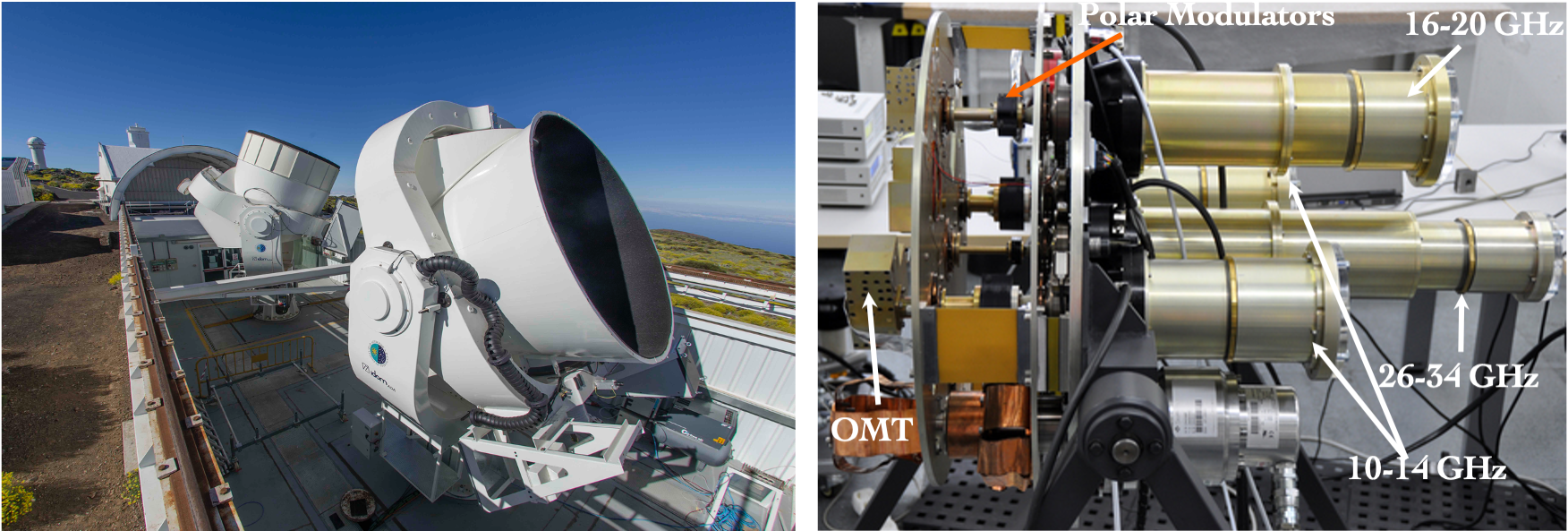}
    \caption{Left: QUIJOTE QT1 and QT2 telescopes, installed in 2011 and 2014 respectively. Both telescopes have a  crossed-Dragone design where the primary mirrors are parabolic with a size of 2.25\,m and the secondary mirrors are hyperbolic with a size of 1.89\,m. Right: the MFI instrument mounted on QT1 (2008-2012).}
    \label{QUIJOTE_image}
\end{figure}

QUIJOTE \cite{Ru} is a CMB experiment located at the Teide Observatory (Tenerife, Spain), a site at an elevation of 2.400\,m with a long tradition of CMB research due to its excellent atmospheric conditions. QUIJOTE is a scientific collaboration between Spain and the UK consisting of two telescopes (cf. Figure \ref{QUIJOTE_image} left). Both telescopes are mounted on a platform that can rotate around the vertical axis at a maximum frequency of 6\,rpm (i.e., 36\,deg/s).

The scientific goals of QUIJOTE are 1) to detect the imprint of gravitational B-modes if they have an amplitude greater or equal to $r= 0.05$ and 2) to provide essential information of the polarization of the synchrotron and the anomalous microwave emissions from our Galaxy at low frequencies (10--40\,GHz).

Several instruments are mounted on both telescopes. On QT1, the MFI (cf. Figure \ref{QUIJOTE_image} right) operated between November 2012 and October 2018 and consisted of four polarimeters (horns): two measuring at 11-13\,GHz and two at 17-19\,GHz. The MFI database used to construct the wide survey consists of approximately 10,000 hours of observations, organized into 1,300 data files, each spanning 8 hours. The MFI signal comprises various components, such as the astrophysical signals we aim to extract (CMB and galactic emission), but also the instrumental noise and the atmospheric signal. All these components add up to form the signal we measure \cite{Ru,Ge}.

%Current 2024, the new version of the MFI, the Second Multi-Frequency Instrument (MFI2) consisting of three polarimeters measuring at 10-15 GHz and two measuring at 15-20 GHz will be installed, and is expected to be 2–3 times more sensitive than MFI. 
%On QT2, the Thirty-GHz Instrument (TGI) and the Forty-GHz Instrument (FGI), consisting both of 31 polarimeters, measure the sky in the 30 and 40 GHz band pass respectively.

%\subsection{MFI database}

% Within these datasets, two horns are operating at 17 and 19 GHz, while the other two horns operate at 11 and 13 GHz.

\section{Atmospheric signal}

The biggest challenge for ground-based telescopes is that they observe through the atmosphere. Even though the atmosphere is in principle unpolarized, CMB polarization observations are affected through leakage of intensity to polarization. Indeed, atmospheric emission is a dominant source of noise in intensity measurements, since the atmosphere emits at CMB relevant frequencies. The primary sources of contamination for the CMB signal are the emissions from molecular oxygen at 60 and 120\,GHz, and the emissions from water vapour at 22 and 183\,GHz \cite{Pa}. Among those, water vapour is the most problematic since it has an inhomogeneous and turbulent distribution, which results in irregular, chaotic and ultimately unpredictable fluctuations of the atmospheric signal \cite{Mo}.\\ \indent
Therefore we need to monitor the Precipitable Water Vapour (PWV) on site in order to properly model the level of contamination and atmospheric extinction at the observing frequencies. 
%that could in turn be used to correct the atmospheric signal and the atmospheric extinction at higher frequencies. 
This monitoring is done by a nearby GPS antenna at Teide Observatory, from which we find a median PWV of 3.4\,mm across all QUIJOTE MFI nominal data observations. In comparison, the median PWV on the Atacama Cosmology Telescope (ACT) site (Atacama Desert, north of Chile) is 2.39\,mm in summer and 1.04\,mm in winter \cite{Mo}. \\ \indent
A useful approach for modeling the inhomogeneous (i.e. turbulent) atmospheric emission is the Kolmogorov model, which postulates that an unconstrained, minimally viscous fluid such as the atmosphere is maximally turbulent and has a velocity field with scale-invariant statistics. According to this model, water vapour will be distributed in agreement with the spatial power spectrum

\begin{equation}
    P(\boldsymbol{k}) \propto |\boldsymbol{k}|^{-11/3},
\end{equation}

\noindent
with $\boldsymbol{k}$ the  wavenumber measuring the spatial frequency of turbulent eddies in the fluid flow \cite{Ko}.\\ \indent
One key ingredient to model the atmospheric signal is the scanning strategy. For QUIJOTE, this strategy consists of  measuring the sky at a constant elevation and in azimuth circles, with a scan speed of 12\,deg/s (one azimuth circle every 30\,s) \cite{Ru}. Given that two horns (2 and 4) of MFI observe the sky at the same frequency band (17--19\,GHz) but in slightly different sky regions (they are separated approximately by $5.6^\circ$ on the sky), they will observe a common-mode atmospheric signal. However, their noise contribution will be statistically independent since each horn is equipped with a different Low Noise Amplifier (LNA). Hence, comparing the signals from those two horns allows the identification of the common atmospheric signal.

%Maybe it is obvious to the specialized reader, but I think that it is worth commenting somewhere that the PWV information provided by the GPS antenna gives the PWV along a certain direction. In this work we are though mostly interested on the study of the spatial and temporal turbulence of the atmosphere, that causes anisotropy on the PWV distribution and then introduces an important noise component on the data as the telescope scans the sky. Even with very high PWV if the WV was perfectly homogeneous, as the O2, its impact would result in an increase of the baseline level, and increase of the thermal noise, but would not contribute to 1/f noise. As I said, this is clear to us and to the specialized reader, but still I think it is worth to making it clear in the text.

\section{Analysis and results}

\begin{figure}[b!]
    \centering
    \includegraphics[width=0.8\textwidth]{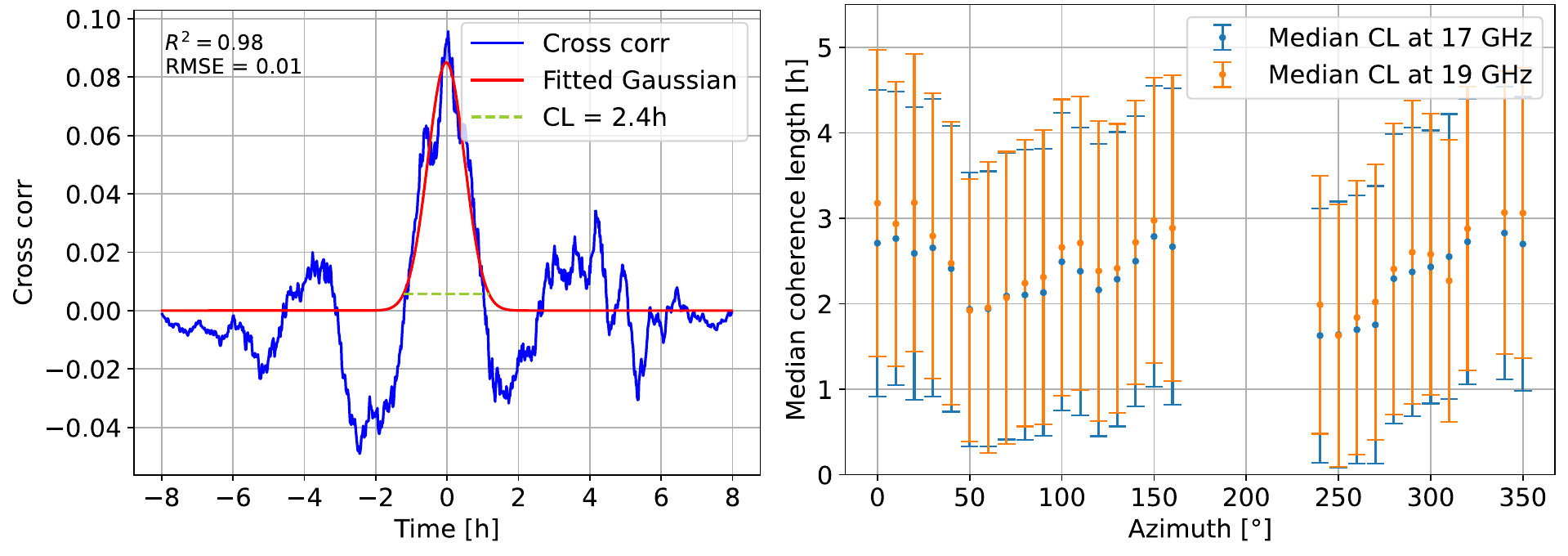}
    \caption{Left: cross correlation function of signal of horn 2 and 4 at 17\,GHz at azimuth 0\,° for one dataset  of QUIJOTE MFI of 8 hours. Right: Median coherence length as a function of azimuth for the whole data of QUIJOTE MFI.}
    \label{fig:analysis02}
\end{figure}

%histogram showing the distribution of coherence length results at 17 GHz (blue) and 19 GHz (purple) obtained at azimuth 0° for all MFI data.

We performed a statistical analysis of the intensity signals from MFI horns 2 and 4 in order to 1) assess the temporal stability of the atmosphere signal and 2) analyze the frequency distribution in the atmospheric signal through the computation of the power spectral density.

First, we computed the cross-correlation function of the signals of horn 2 and 4 at 17\,GHz and then at 19\,GHz. The discrete cross-correlation function between two signals $f$ and $g$ at time shift $\tau$ is given by

\begin{equation}
    C[\tau] = \frac{\sum_{n=0}^{N-1} \overline{f(n)} g(n + \tau)}{N \sqrt{\sigma_f^2 \cdot \sigma ^2_g}},
\end{equation}

\noindent
where $N$ is the number of elements in both signals $f$ and $g$, and $\sigma_{f/g}^2$ is the variance (auto-correlation function at zero lag). After selecting all data with a given azimuth value, the correlation function is computed for all possible time shifts and plotted against this time shift (cf. Figure \ref{fig:analysis02} left for an example with one data file of MFI). A maximum in the correlation function implies a strong similarity of the signals, while a correlation curve close to zero indicates no correlation. On those plots, we systematically observe a correlation peak at zero lag. This peak indicates that both horns measure the same atmospheric signal.

In order to quantify the time coherence of the correlation between the two signals, we define the Coherence Length (CL) as the characteristic time over which the signals remain correlated, reflecting the duration of stability in the atmospheric signal. The coherence length was determined from the width of the correlation peak by fitting a Gaussian curve to each dataset's correlation peak and determining the width of the Gaussian curve at $1/15$ of the Gaussian maximum amplitude. Mathematically, this is expressed as $CL = 2  \sqrt{2  \ln(15)}\sigma$ with $\sigma$ the standard deviation of the Gaussian fit. After calculating this coherence length for each dataset of MFI at each azimuth (in order to compare the same area of the sky), we derived the median coherence length at each azimuth (cf. Figure \ref{fig:analysis02} right). Our findings indicate that the atmosphere stays stable for a duration of approximately 2 to 3 hours.

\begin{figure}[t]
    \centering
    \includegraphics[width=\linewidth]{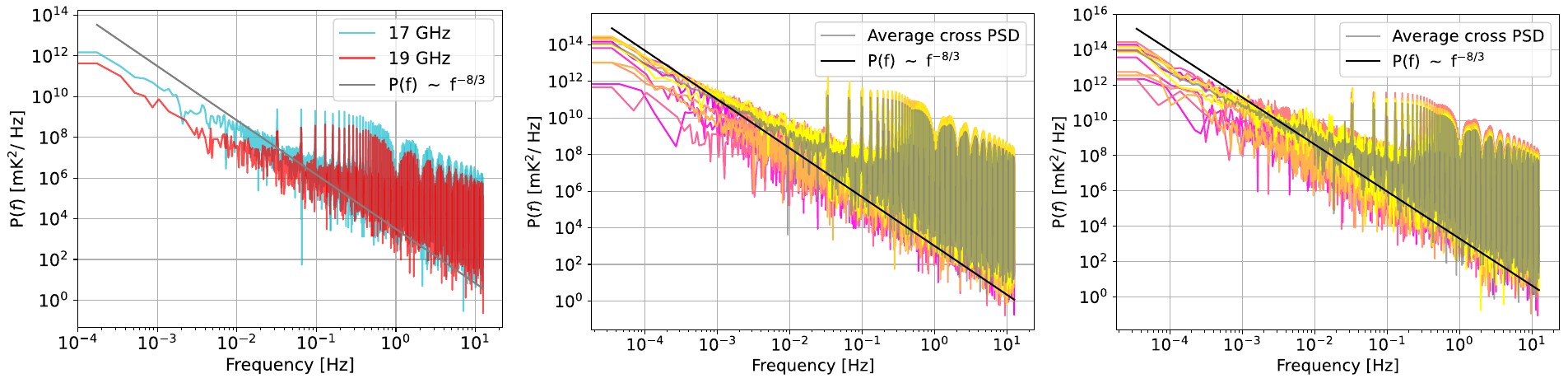}
    \caption{Left: cross power spectral density (CPSD) of the signal of horn 2 and 4 at 17\,GHz (blue) and 19 GHz (red) for one dataset, along with a $-8/3$ power law fit. Middle: CPSD of several datasets (pink to yellow gradient) with average CPSD (gray) at 17\,GHz for a telescope elevation of $30^\circ$.  Right: the same at 19\,GHz.}
    \label{fig:PSD}
\end{figure}

Furthermore, it is noteworthy to verify whether our atmospheric signal follows a Kolmogorov spectrum. In order to do so, we computed the cross power spectral density (CPSD), which represents the distribution of a signal across different frequencies in the frequency spectrum, for signals of horn 2 and 4 at 17\,GHz and then at 19\,GHz (cf. Figure \ref{fig:PSD}). The CPSD is defined as

\begin{equation}
    \Gamma_{fg} = \Re{\big ( \hat{f} \cdot \overline{\hat{g}}\big )},
\end{equation}
\noindent
with $\hat{f}/\hat{g}$ the Fourier transform of signal $f/g$. We found that the CPSD of our data follows roughly a Kolmogorov spectrum.

\section{Conclusion and outlook}

The atmosphere contaminates astrophysical signal at CMB relevant frequency for ground based telescopes. Being able to model and extract the atmospheric signal correctly is essential in order to reach the sensitivity needed to measure the B-modes. In this analysis, we found that the atmospheric signal remains stable over a period of 2 to 3 hours, and that the water vapour is distributed roughly according to a Kolmogorov spectrum. Next, we will investigate the relation between wind speed and coherence length, and the variation of the atmospheric signal within a year.

\end{document}